\documentclass[letter]{aa} 

\usepackage{graphicx}
\usepackage{natbib}
\usepackage{float}
\usepackage{color}
\usepackage{txfonts}
\usepackage[english]{babel}
\usepackage{gensymb}
\begin{document} 

\newcommand{\pone}{P_1}
\newcommand{\ptwo}{P_2}
\newcommand{\pthr}{P_3}
\newcommand{\hpthr}{\hat{P}_3}
\newcommand{\src}{PSR B0943+10}

\title{The Discovery of Transitive Phenomenon in the Radio Emission of the mode-switcher PSR B0943+10}

   \author{S.~A.~Suleymanova
          \inst{\ref{prao}}                             
          \and
   A.~V.~Bilous\inst{\ref{astron}}  }

    \institute{P. N. Lebedev Physical Institute of the Russian Academy of Sciences, Pushchino Radio Astronomy Observatory, Pushchino 142290, Russian Federation\label{prao}
    \and
    ASTRON, the Netherlands Institute for Radio Astronomy, Postbus 2, 7990 AA Dwingeloo, The Netherlands\label{astron}}

\abstract{\src\ is known to switch between two distinct, hours-long modes of radio emission, Bright (B) and Quiet (Q). Up to now the switches in both directions were believed to occur instantly (on the scale of a spin period). We have found a transitive process around the Q-to-B-mode switch, which consists of two additional short-lived modes, each with distinct average profiles and subpulse drift rates. Based on observations at low radio frequencies, we examine the properties of these transitive modes and discuss their implications in the framework of the traditional carousel model of drifting subpulses.
}
   

\titlerunning{Transitive Phenomenon in the Radio Emission of B0943+10}
\maketitle

%

\section{Introduction}
\label{sec:intro}
\src\ is a classical example of the so-called ``mode-switching'' pulsars, a small subset  
of radio pulsars that switch between several stable states of the electromagnetic emission. 
For \src, two modes have been identified so far: "Bright" (or "Burst") and ``Quiet'' \citep[hereafter B-mode and Q-mode, see][]{Suleymanova1984}.
In radio band, both the average pulse shape and the properties of single-pulse emission change between the modes, with 
B-mode being notorious for the organized temporal drift of individual subpulses \citep[e.g.][]{Deshpande2000}. 

In the X-ray band, the radio mode switch is accompanied by changes in average pulse morphology and flux density \citep{Hermsen2013,Mereghetti2016}. The X-ray emission in both modes has a pulsed thermal component 
originating in the hot polar cap region on the neutron star surface, few hundred kilometers away from the presumable origin of the magnetospheric radio emission \citep{Bilous2014}. Thus, some global-scale magnetospheric transformation is expected during mode transitions \citep{Timokhin2010,Cordes2013}.

Studying the properties of mode transitions may help constraining the nature of said transformation
and the searches for peculiar single-pulse behavior around mode switches have been conducted regularly. To the extent of authors' knowledge, no such effects have been found so far: recorded mode transitions happen instantly and abruptly \citep{Bartel1982,Wang2007,Esamdin2005,Rajwade2021}. 

For \src\, mode transitions have been considered to be instantaneous as well, although there is no clear corroboration of this in the published literature.
In this paper we report the discovery of transitive phenomena around Q-to-B mode switches in several archival low-frequency observations  conducted  at 112\,MHz with Large Phased Array (LPA) at Pushchino Radio Observatory (PRAO), and at 25$-$80\,MHz with the LOw-Frequency ARray \citep[LOFAR,][]{vanHaarlem2013}. This transitive state (hereafter T-mode) lasts for about 1 minute or 60 pulsar rotations, thus being much shorter than the hours-long main modes. We describe the peculiar single-pulse properties during 
T-mode and speculate on their implications for the mode switching theories.

\section{Observations and data processing}
\subsection{PRAO}

LPA is a transit instrument and the duration of observing session can not exceed 3.5\,min/$\cos\delta$, where $\delta$ is the declination of the source. For \src, with its spin period $\pone=1.0977$\,s, this translates to a maximum of 194 pulses available per session. The pulses were recorded using a digital receiver with a bandwidth of 2.5\,MHz divided into 512 4.88-kHz spectral channels by a Fast Fourier Transform (FFT) processor. The actual total bandwidth taken into processing was 2.245\,MHz. The signal was dedispersed to the frequency of 111.88\,MHz (hereafter 112\,MHz). 

The narrow bandwidth of the channels reduced the effect of pulse broadening caused by interstellar dispersion delay. At 112\,MHz the total pulse delay within a single channel for \src's DM of 15.4\,pc\,cm$^{-3}$ is 0.5\,ms, much smaller than the time resolution of 2.8672\,ms. More detailed description of LPA's digital receiver and data pre-processing is given in \cite{Suleymanova2012}.

With a rotation measure of $15\pm1$\,rad/m$^{-2}$ \citep{Suleymanova1998}, the Faraday modulation period at 112\,MHz is 
about 1.6\,MHz,  which is comparable to the total bandwidth of the receiver. Thus, 
despite LPA being a linearly polarized array, the recordings provide a reasonable estimate of the pulsar's total intensity emission 
\citep[see also][]{Suleymanova2009}.

\subsection{LOFAR}
\src\ was observed with the low-band antennas (LBAs) of LOFAR core stations at a centre frequency of 53.8\,MHz,
and 25600 channels across a 78.1-MHz bandwidth. The time resolution of the raw data was 0.65\,ms. Observations were pre-processed with the standard LOFAR pulsar pipeline \citep{Stappers2011}.
The signal was dedispersed and folded synchronously with the topocentric pulsar period obtained with the ephemerides from \citet{Shabanova2013}. The dedispersion was subsequently refined using a more precise value of the dispersion measure (DM), obtained by measuring the $\nu^{-2}$ lag of the fiducial point in the B-mode profile of the same observations \citep{Bilous2014}. 
Each period contained 512 longitude bins, corresponding to time resolution of about 2\,ms. 
For further details of data acquisition and pre-processing we refer the reader to \citet{Bilous2014} and \citet{Bilous2018}. 

For both LOFAR and PRAO data the subsequent analysis was performed on frequency-integrated pulse stacks, 2D arrays of the
uncalibrated total intensity as a function of rotational longitude $\phi$
and time $t$, $s(\phi, t)$. In this notation, time is assumed to be 
constant within each rotational period of pulse stack: $t=[$pulse number$]\times\pone$.

\section{Overview of the radio emission in main B- and Q-modes}

\subsection{Single pulses}

\begin{figure}
\centering
 \includegraphics[width=0.4\textwidth]{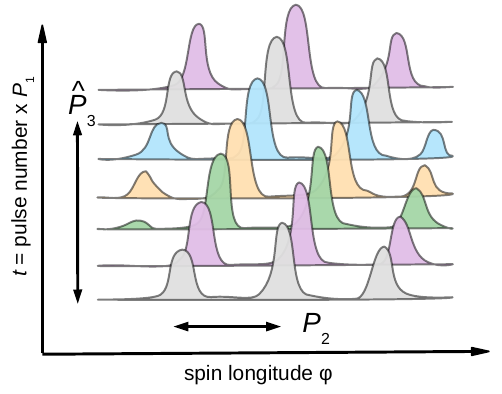}
 \caption{Cartoon example of a drifting subpulse sequence, characterized by the 
longitudinal spacing between subpulses  $\ptwo$ and the amount of time between successive appearances at the same 
spin longitude $\hpthr$. In this example subpulses are drifting towards the trailing edge of the onpulse window. For clarity, here we draw $\hpthr$ being equal to the integer number of rotation periods, 
however this is not necessarily true in general.} 
 \label{fig:drfit_cartoon}
\end{figure}

Soon after the discovery of \src\ in 1968 at PRAO  \citep{Vitkevich1969}, a remarkable organized change of its subpulse positions had been noticed by \citet{Taylor1971}. Further investigations at 430\,MHz demonstrated that subpulses shift within the onpulse window with the rate of approximately $4\fdg4$  per stellar rotation \citep{Backer1975}.

Following \citet{Edwards2003} and \citet{Bilous2018}, the subpulse position $\mathrm{pos}_\mathrm{drift}$ can be mathematically expressed as follows:
\begin{equation}
 \mathrm{pos}_\mathrm{drift}(\phi, t) \sim \mathrm{Re}\left[\exp \left(i\dfrac{2\pi}{\hpthr}t + i\dfrac{\pone}{\ptwo}\phi\right)\right].
 \label{eq:sdrift}
\end{equation}
Here $\hpthr$ is the observed modulation period along the lines of constant $\phi$ and $\ptwo$ is the longitudinal spacing between subpulses recorded within same pulse period (see Fig.~\ref{fig:drfit_cartoon}).

Both $\hpthr$ and $\ptwo$
can be positive or negative, and the apparent direction of drift (i.e. whether pulses seem to march towards the leading or trailing edges of the profile) depends on whether $\pthr$ and $\ptwo$ have the same sign. 
Following \citet{Edwards2003},
we assume that $\ptwo$ is always positive and let the sense of drift be determined by the sign of $\hpthr$.

$\hpthr$ is usually determined from the longitude-resolved fluctuation spectra (LRFS). LRFS consist of 1D Fourier transforms of the pulse stacks, computed over the 
lines of constant rotational longitude $\phi$. When drift is present, LRFS have two peaks at Fourier 
frequencies of $\nu_t=\pm 1/\hat{P}_3$. In this case 
$\ptwo$ can be assessed from evolution of the complex phase of the Fourier transform with the longitude. If $S(\phi, \nu_t)$ is the longitude-resolved Fourier 
transform of $s(\phi,t)$ computed over the lines of constant $\phi$, then from Eq.~\ref{eq:sdrift} it follows that
\begin{equation}
\begin{aligned}
&\mathrm{Arg}\left[ S(\phi,\nu_t =1/\hpthr)\right]  \sim \ptwo\phi, \\
&\mathrm{Arg}\left[ S(\phi,\nu_t =-1/\hpthr)\right]  \sim -\ptwo\phi. 
\end{aligned} 
\label{eq:LRF_peaks}
\end{equation}
Selecting the peak with the observed positive phase gradient will fix the sign of $\hpthr$.

\begin{figure*}
\centering
 \includegraphics[width=0.33\textwidth]{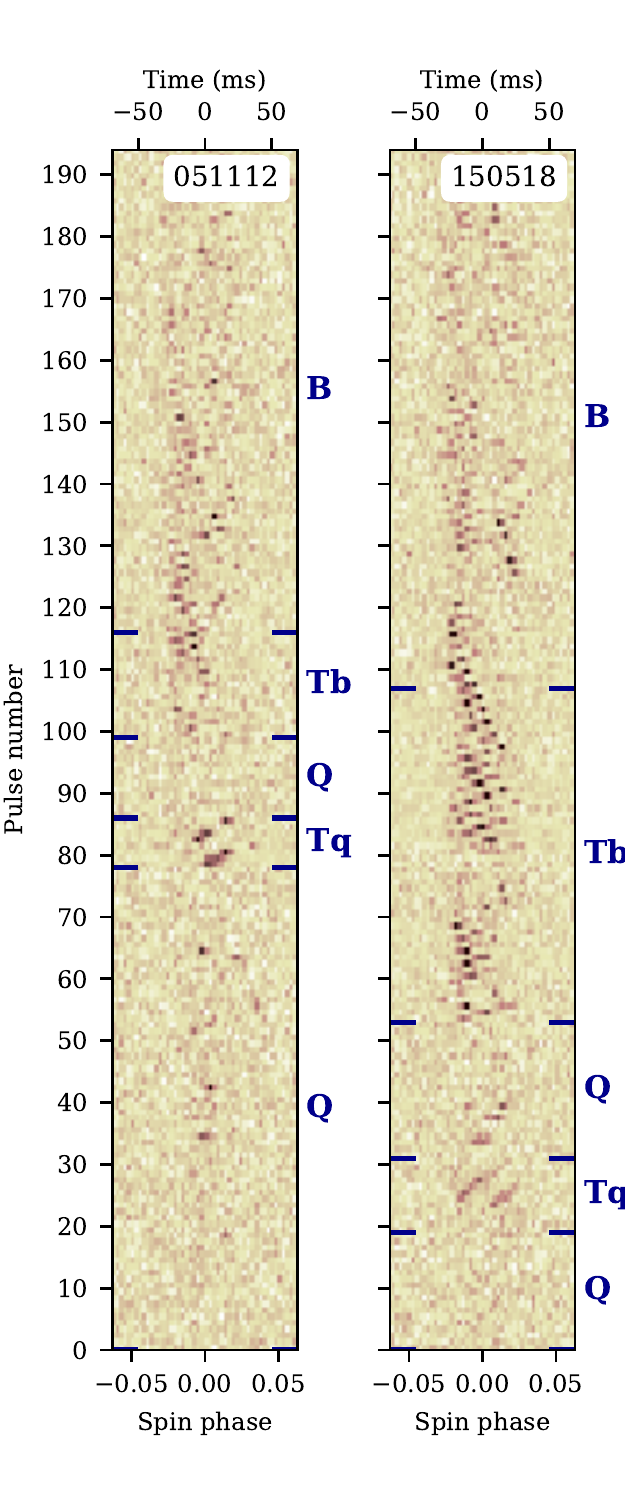}\includegraphics[width=0.33\textwidth]{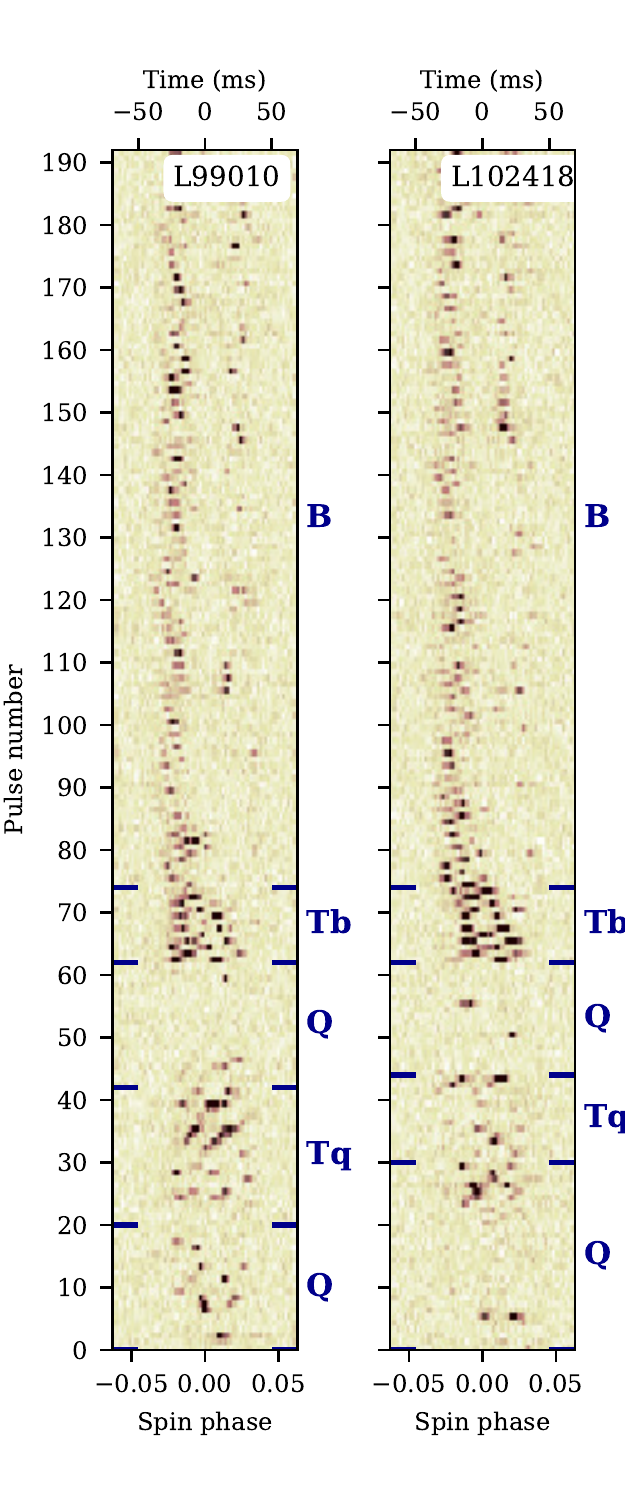}\includegraphics[width=0.33\textwidth]{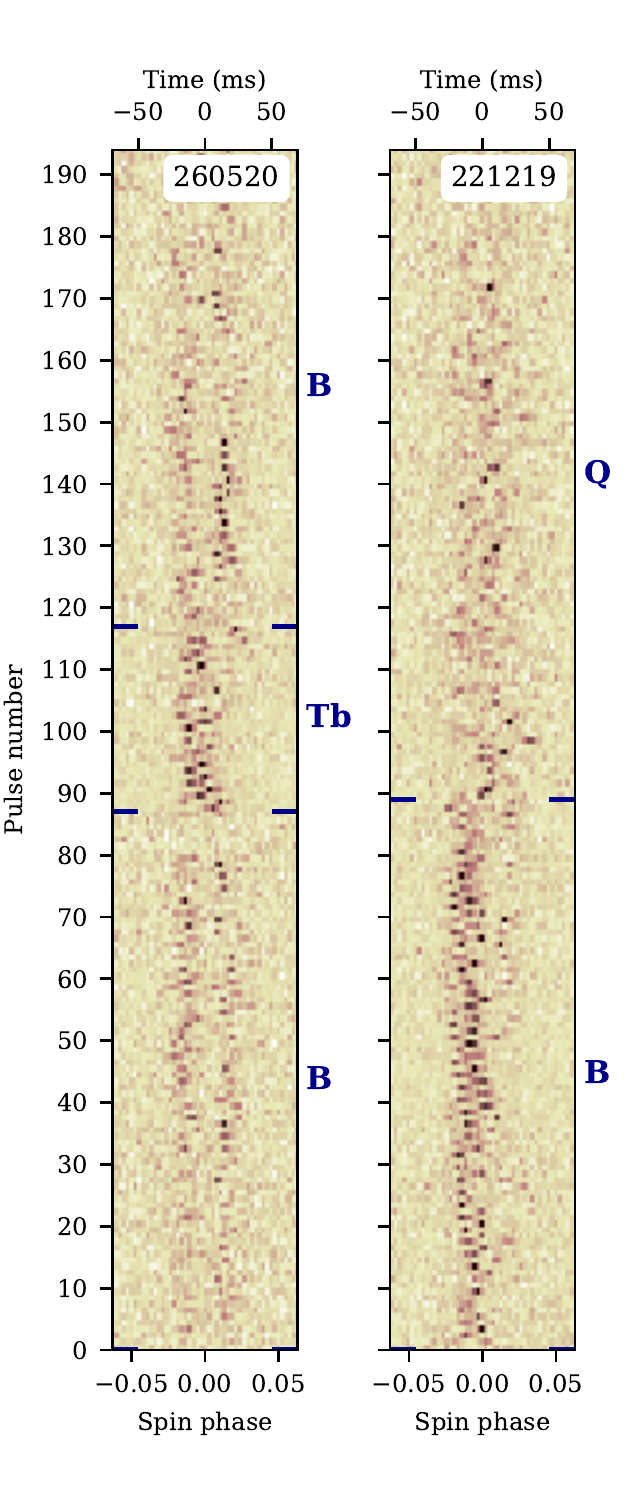}
 \caption{Single-pulse sequences for six observing sessions at LPA (112\,MHz) and LOFAR (30$-$80\,MHz) telescopes. The designations for sessions at LPA are given in \textit{ddmmyy} format. Observations with IDs starting with L were taken at LOFAR. For all sessions only the on-pulse region is shown and the colors are saturated to highlight fainter pulses. Mode boundaries were determined based on the average profile morphology and the presence of subpulse drift in 2D FFT spectra (Sect.~\ref{sssubseq:TbTq}). Sessions 051112, 150518, L99010 and L102418 feature Q-to-B mode transitions.  Session 260520 shows a rare B-Tb-B mode transition and 221219 exhibits a typical featureless B-to-Q mode switch.} 
 \label{fig:waterfall}
\end{figure*}

For \src, $\hpthr \approx -2.2\pone$ and $\ptwo=11\degree$, meaning 
that subpulses drift to the trailing edge of the onpulse window \citep{Bilous2018}. However, 
the absolute value of $\hpthr$ is commonly used in the literature \citep[e.g.][and other papers in this series]{Deshpande2000}. 
Since $\hpthr$ is close to a small multiple of $\pone$, the apparent drift is not obvious: the subpulses 
exhibit even-odd modulation together with a slow shift towards the leading edge of the average profile (see Fig.~\ref{fig:waterfall}).  

Measuring $\hpthr$ with LRFS does not solve the problem of aliasing. This problem is caused by the undersampling of 
the subpulse motion (subpulses can be observed only for a small fraction of pulse period) and by volatile subpulse shapes, which prevent tracing the path of each specific subpulse throughout a pulse sequence \citep{vanLeeuwen2003}.
The true drift period $\pthr$ in presence of aliasing is related to observed $\hpthr$ as 
\begin{equation}
 \dfrac{\pone}{\pthr} = n +  \dfrac{\pone}{\hpthr},
\end{equation}
with the signed integer $n$ being the degree of aliasing. Note that the true direction of drift rate can be different from 
the observed one. 

In the work of \citet{Deshpande2000}, $n$ was determined to be 1
based on fleeting $37\pone$ modulation of subpulse intensity. This modulation was interpreted within the popular carousel model 
\citep{Ruderman1975}. In this model, pulsar emission comes from discrete spots located on a ring centered on 
magnetic pole. The slow rotation of the carousel causes the observed subpulse drift. \citeauthor{Deshpande2000} argued that in their observations the individual subpulses retained their relative distribution of amplitudes for approximately $256\pone$, 
which allowed to trace individual subpulses and resolve aliasing. Such amplitude modulation is extremely rare, with only two other 
instances being found despite extensive searches \citep{Backus2011,Bilous2018}. 

\src\ is unique in its slow consistent evolution of the drift properties. \cite{Rankin2006b} have found that \src\'s subpulse drift rate changes exponentially by some 5\% during several hours after B-mode onset with the characteristic time of about 73 min. The change is faster at B-mode onset. Later it was found that this behavior is independent of observing frequencies between 40 and 327\,MHz and over the timescale of several years \citep{Rankin2006b, Backus2011, Suleymanova2017, Bilous2018, Suleymanova2021}. 

In turn, emission in the Q-mode is stable in the sense that no gradual changes in its properties have been recorded. Despite systematic searches, no drifting subpulses have been detected in the Q-mode \citep{Suleymanova1984,Backus2010}, except for 
an occasional feature at 0.0275\,cycles/$\pone$ corresponding to $\hpthr = 36.4\pone$ \citep{Rankin2006b}. The origin of this feature is unclear and no similar features 
have been found since then \citep{Bilous2018}.

\subsection{Average profile}

In both modes the average pulse profile is composed of two components of equal width which are moving away from each other at lower frequencies. The full width at half maximum of the components is about $7\fdg2$ for the B-mode and $11\fdg5$ in Q-mode and does not show a strong dependence on frequency between 40 and 80\,MHz \citep{Bilous2014}. Recently \cite{Suleymanova2021} have shown that the width of the average pulse components in the B-mode  is constant over the wider frequency range  of 62$-$1391\,MHz with the separation between components increasing rapidly towards lower frequencies in accordance with a power law: 
\begin{equation*}
s(^{\circ})= 130\fdg8 \times \nu_\mathrm{MHz}^{-0.56}.
\label{formula}
\end{equation*}

In B-mode the leading component is generally brighter than the trailing one,
however the ratio of components' peak amplitudes changes in a frequency-dependent manner as the mode evolves. For the first few minutes of the B-mode, the trailing component may be brighter, again in a frequency-dependent manner.
The fiducial longitude (midpoint between profile components, corresponding to the moment of time 
when the line-of-sight (LOS) passes closest to the magnetic axis) is systematically delayed by about $1\fdg4$ 
per mode, resetting at the next Q-to-B transition. The amplitude of delay does not depend on frequency at least between 40 and 327\,MHz \citep{Bilous2014,Suleymanova2021}. No such effect has been identified for Q-mode.

\section{The transitive phenomenon around Q-to-B transition}

\subsection{Identification of the Tb- and Tq-modes}

\begin{figure}
\centering
 \includegraphics[width=0.45\textwidth]{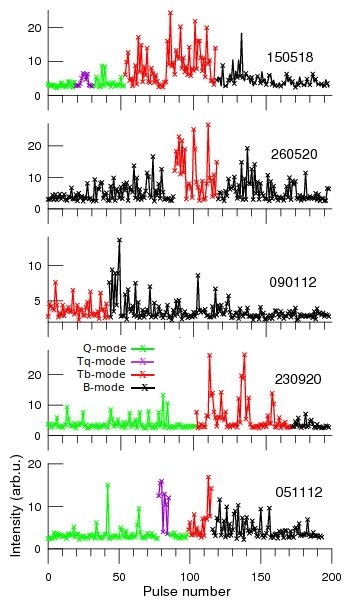}
 \caption{Five records of Q-to-B mode switches from the archives of 112-MHz observations. The intensity of subpulses in units of signal-to-noise ratio is given as a function of pulse number. Individual modes are highlighted with different colors (black for B-mode, green for Q-mode, red for Tb-mode, and magenta for Tq-mode).} 
 \label{fig:fig2}
 \end{figure}

Figure~\ref{fig:waterfall} shows examples of individual pulse sequences for some of our observations. In general, the difference between the B- and Q-modes is evident to the naked eye. In the B-mode, individual pulses form two columns corresponding to components of the average profile. The drift of subpulses near central longitudes is not visible due to a decrease in average pulse intensity in the saddle between the components. In the Q-mode subpulses are distributed randomly within the on-pulse window.

Panels 150518, L99010, L102418 and 260520 show unusual behavior of subpulses around Q-to-B mode transitions. 
When subpulse drift recommences after the end of Q-mode, for a short period of time subpulses drift through
entire on-pulse window and, unlike later on, there is no fading of subpulses at central longitudes. During this period
the on-pulse window is somewhat narrower than in well-established B-mode sequences.

These features allowed us to identify these particular sequences of pulses as a new mode, which we designated ``Transitive B-like'' mode, or Tb-modes. 
Tb-mode has been recorded in all available observations except for session L169237 in \citet{Bilous2014}. 
Comparing with the sessions presented on Fig.~\ref{fig:waterfall}, the S/N ratio of the pulsed signal is smaller there, 
and emission is modulated by scintillation, with one of the intensity troughs coinciding with Q-to-B transition
\citep[see Fig.~1 in ][]{Bilous2014}.

On panel 260520, the subpulse drift pattern characteristic to Tb-mode is observed in the 30-second interval between B-mode pulses. There are no Q-mode pulses in this record, but the shape of the average B-mode pulse profile, with two equally bright components, indicates that the Q-to-B switch occurred a few minutes before the start of observing session \citep{Rankin2006b, Suleymanova2009}. The presence of Tb-mode pulses within the B-mode sequence suggests that B-mode emission in the first minutes after mode onset is unstable.

The Tb-mode is not the only transitive phenomenon around Q-to-B switches. Some time before Q-mode cessation 
there appear groups of pulses that slowly drift towards the trailing edge of the on-pulse window. This drift is present in all but 
one Q-to-B transitions observed. 
We have labeled these slowly drifting pulses as Transitive Q-like modes, or Tq-modes. 

\begin{figure*}
\centering
\includegraphics[width=0.5\textwidth]{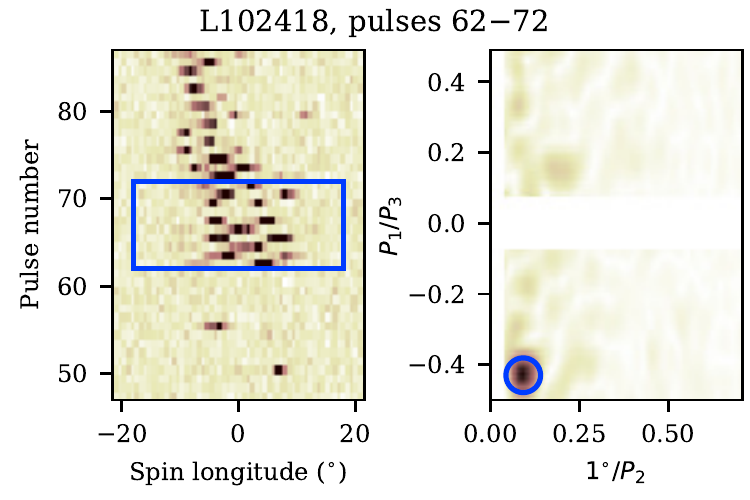}\includegraphics[width=0.5\textwidth]{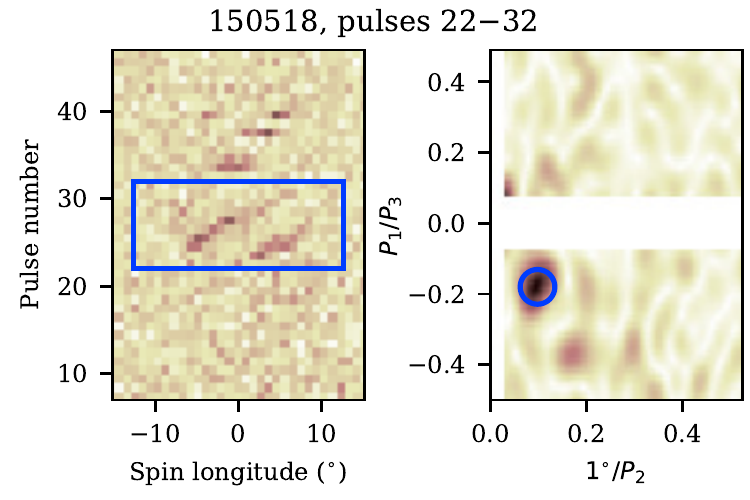}
  \caption{An example of a 2D (spin longitude vs pulse number) FFT spectrum featuring the Tb-mode of session L102418 (left) and Tq-mode of session 150518 (right). The Fourier transform was taken from the region enclosed in the blue rectangle on the left-side pulse sequence plot. Frequencies close to zero were eliminated from analysis to reduce the influence of red noise on the peak finding algorithm. 
  The colorbar range in each figure is set by the local minimum and maximum and is not uniform across different LRF spectra.}
\label{fig:FFT_Tb}
\end{figure*}

Both Tb- and Tq-modes exhibit drifting subpulses and the presence of this drift was used to establish mode boundaries. However, the 
Tb-mode does not follow Tq-mode immediately. Between them there is a gap of several spin periods with very scarce single pulses
resembling those of the Q-mode. Interestingly enough, the only observed B-Tb-B transition also exhibits the absence of emission for
a few seconds before the Tb-mode starts (Fig.~\ref{fig:waterfall}, panel 260520).

Because the transit time of \src\ through the BSA beam is much smaller than the typical duration of its main modes, the records of 
mode switching are rare,  
with only five sessions with Q-to-B transitions being found in the archives. Subpulse intensities for
these five records are shown in Fig.~\ref{fig:fig2}, with different modes shown by different colors.
There are no Q-mode pulses in the 090112 session. The low values of the subpulse drift rate indicated that the first 42 pulses correspond to the Tb-mode, and the remaining pulses to the B-mode. In addition, the shape of the averaged pulse in B-mode with the ratio of components $R(2/1)=2.1$ clearly pointed out to the Q-to-B switch occurring shortly before the start of the observation session \citep{Rankin2006b, Suleymanova2009}.

In five cases on Fig.~\ref{fig:fig2} the Tq-mode
is much shorter than the Tb-mode, and in one session (230920) it is absent completely. The duration of the Tb-mode varies between sessions, comprising 17$-$66 spin periods or 19$-$69\,s. At lower frequencies, the two transitions recorded exhibit shorter Tb-modes -- only 13\,s, however this may be just a coincidence. More low-frequency observations are needed to explore the Tb-mode duration below 100\,MHz.

To our surprise, the opposite, B-to-Q mode transitions were featureless (e.g. panel 221219 on Fig.~\ref{fig:waterfall}). No transitive phenomena were found in any of our recordings.

\subsection{Subpulse drift}

\subsubsection{Tb- and Tq-modes}
\label{sssubseq:TbTq}

Both Tb- and Tq-modes are much shorter than pulse sequences typically used for computing LRF spectra and determining drift parameters in the B-mode. Cutting the usual few-hundred pulse sequences around Tb- and Tq- modes pollutes the spectra with either B-mode drift or Q-mode noise, diminishing the S/N of the transitive mode spectra peaks, which is detrimental for the fainter Tq-mode sequences. 
However, finding peaks on LRF spectra usually involves
integrating spectral power within onpulse window, thus not taking into account the regularity of the 
longitude separation between subpulses in the same period (this information is stored in the complex phase of LRF spectra). 
The 2D FFT transform \citep{Edwards2003} does not have this disadvantage and thus is better suited for measuring the drift parameters of the short pulse sequences of the Tq- and Tb-modes.

To measure $\hpthr$ and $\ptwo$ simultaneously, we performed 2D FFT transforms in a fixed-size window sliding along the pulse number axis. The window
size was 10 pulses by $22\degree$ or $40\degree$ of rotational longitude for LPA and LOFAR observations, respectively. The sliding step was $2\pone$. In order to measure the position of FFT peak more precisely, sequences of zeros were added to each data row, boosting the nominal Fourier frequency resolution by a factor of 10. Formal errors on $\hpthr$ and $\ptwo$ were derived from the frequency resolution of the padded data transforms, however the real, noise-influenced errors are supposed to be larger by a factor of a few.

To estimate the significance of recorded features we performed the same feature extraction procedure on 
original pulse sequences but with randomized subpulse positions. The randomization was done by shuffling 
the pulse periods within the groups of 4 phase bins. This way the average shape of 10-pulse integrations was preserved, while destroying any periodicity along the axis of constant longitude and attenuating the 
periodicity in orthogonal direction.
This procedure was performed 1000 times and the percentile of peak on FFT spectrum of the real data was  
compared to the distribution of the corresponding peaks of the simulation. 
Given the total length of the pulse sequences explored, the total number of chance detections above our 
threshold is estimated to be between 0.5 (independent 10-pulse sequences) and 6 (independent 2-pulse sequences).

Examples of 2D FFT spectra are given on Fig.~\ref{fig:FFT_Tb}. Most of the features detected came from Tb- and B-modes where subpulse drift is visible to the naked eye. In all but one sessions
there had been significant detections of drift periodicity in the Q-mode short before the Q-to-B transition (Tq-mode).
In both transitive modes, as well as in the B-mode the feature at positive $\ptwo$ had negative $\hpthr$, 
meaning that pulses drifted to the trailing edge of the onpulse window.

\begin{figure*}
\centering
 \includegraphics[width=0.5\textwidth]{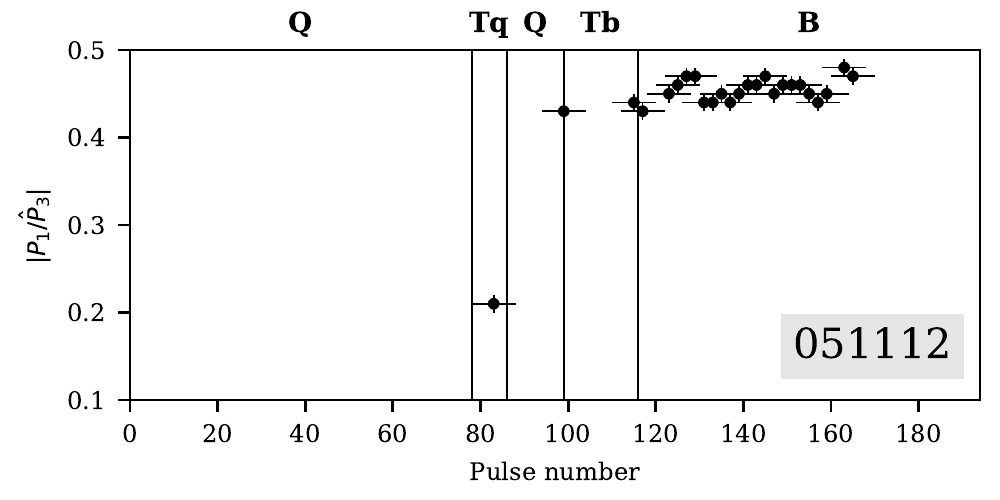}\includegraphics[width=0.5\textwidth]{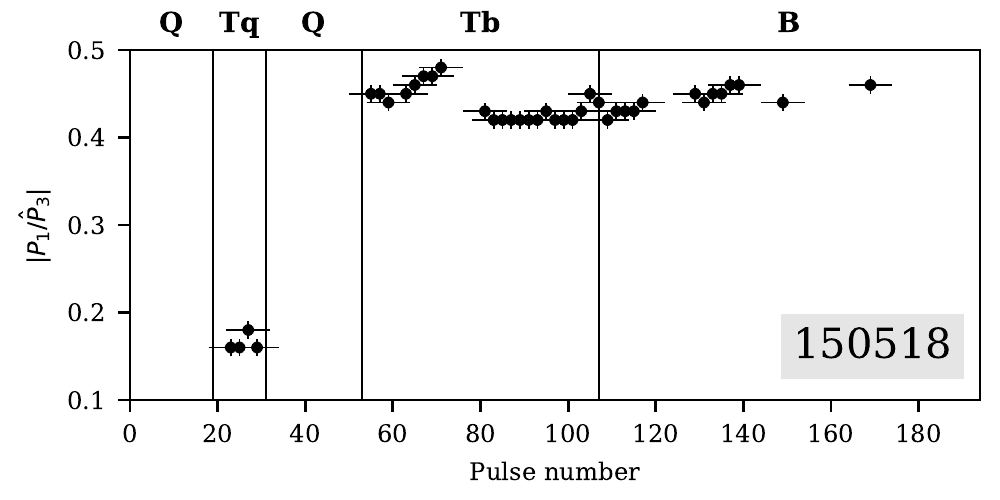}
 \includegraphics[width=0.5\textwidth]{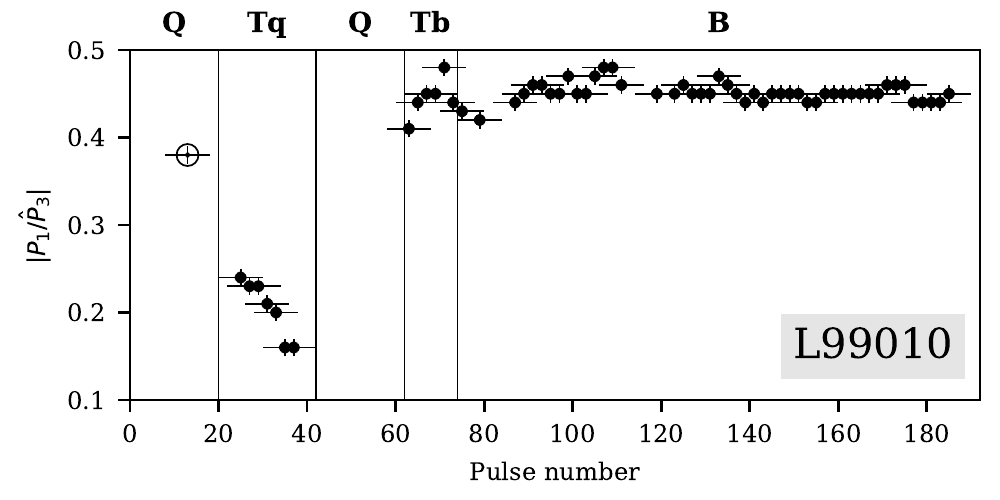}\includegraphics[width=0.5\textwidth]{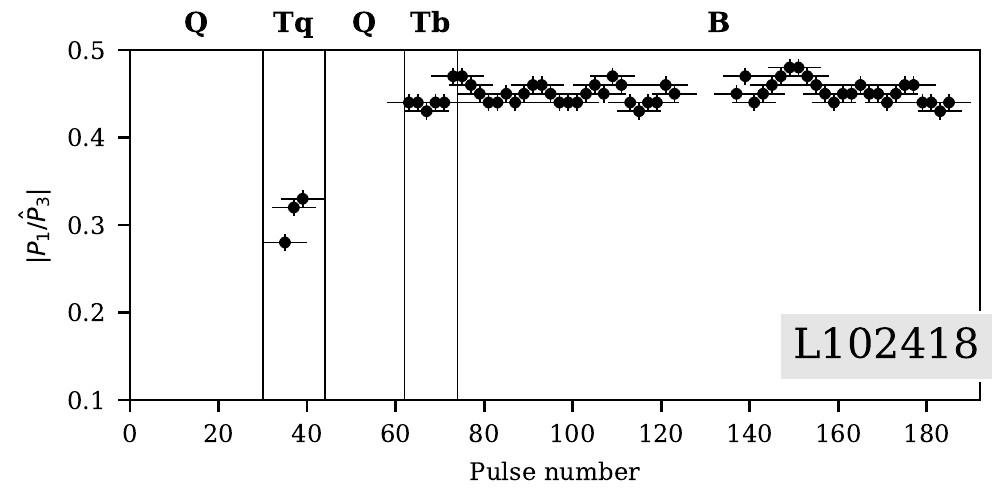}
 \includegraphics[width=0.5\textwidth]{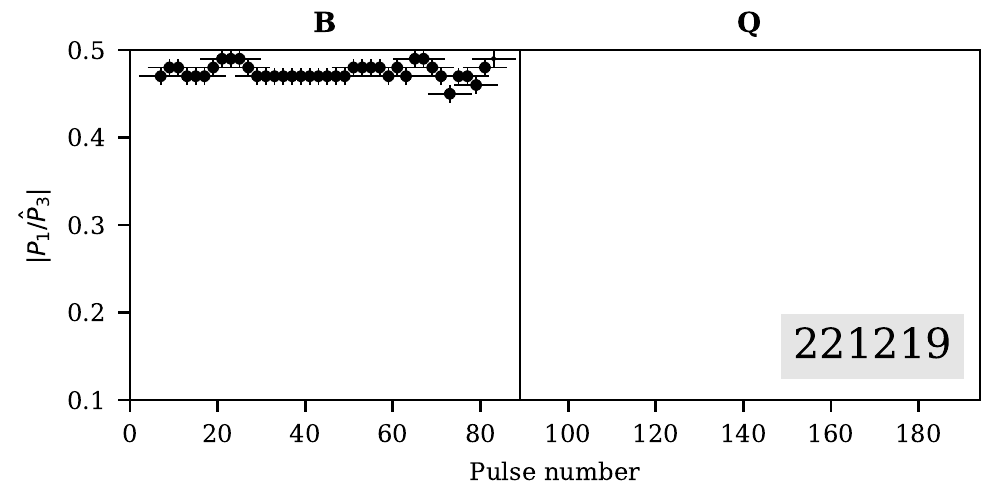}\includegraphics[width=0.5\textwidth]{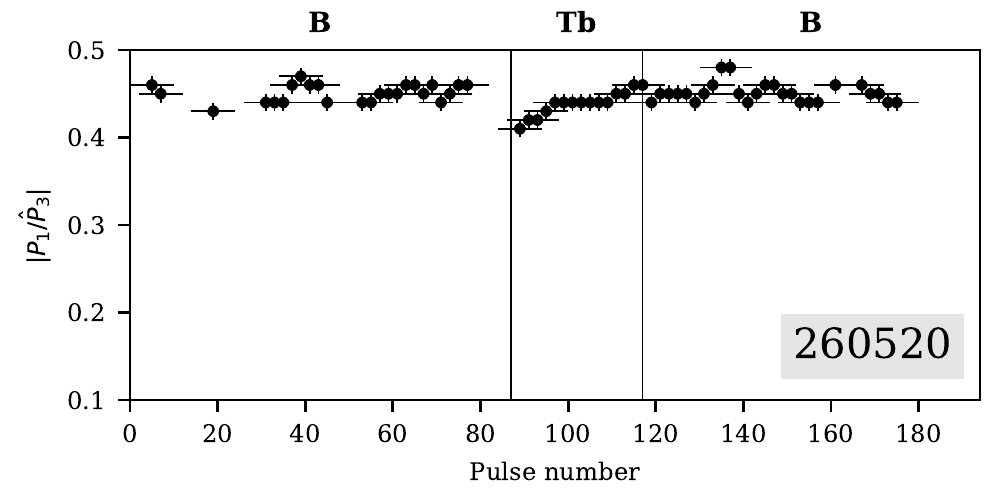}
   \caption{The absolute value of the modulation frequency around the lines of constant spin phase ($f_3\equiv |\pone/\hpthr|$) as a function of time for the pulse sequences from Fig.~\ref{fig:waterfall}. Horizontal errorbars mark the edges of 10-pulse samples used for $f_3$ calculation. 
  Vertical errorbars show the nominal frequency determination error taking into account the zero-padding. 
  For each subsample, only $f_3$ with the highest S/N is plotted (see text for details on S/N estimates). Only
  peaks which were stronger than 99.5\% of the simulated sample are shown. The vertical lines mark mode edges. All points here correspond to negative $\hpthr$ except for the hollow one for session L99010. For the PRAO sessions, the absence of $f_3$ points near the observation end is due to pulsar signal attenuation.}
\label{fig:F3}
\end{figure*}
 
$\hpthr$ in the Tq-mode varies substantially between mode instances, ranging from approximately $3.2\pone$ to $6.7\pone$. It also exhibits variation within the Tq-mode itself. For the Tb-mode $\hpthr$ stays closer to the B-mode values, most of the times being larger than in the B-mode. The jitter of measured $\hpthr$ in the B-mode is partially due to noise influence, and partially intrinsic -- in \citet{Bilous2018} LRF spectra on 512-pulse sequences recorded multiple peaks with frequency spread on the order of $0.01\pone/\pthr$.

Interestingly enough, the subpulse separation remains more or less constant throughout Tq-, Tb- and B-modes,  varying chaotically around $11\degree$ with a magnitude of  $1\degree-2\degree$.
$\ptwo$ for the same LOFAR sessions  was also measured in \citet{Bilous2018} using a more precise phase track method on 512-pulse sequences, which takes into account the observed increase of subpulse separation at the edge of the onpulse window due to the curved LOS path. In their work $\ptwo$ at the fiducial longitude (middle of onpulse window for \src) is $10\fdg6-10\fdg8$, with $\ptwo \approx 18\degree$ at $\pm 15\degree$ of spin longitude. 
In our case, the 2D FFT gathers information from the entire onpulse window, thus our $\ptwo$ is larger. Since pulse 
sequences comprise only 10 pulses, the illumination of the onpulse window is quite uneven, bringing extra 
variability in measured $\ptwo$.

\subsubsection{Comparing drift properties in Tb- and B-modes}

The shape of the average pulse changes abruptly during Tb-to-B transition, however the 
evolution of drift parameters seems to be smooth. It is interesting to compare $f_3\equiv |\pone/\hpthr|$ in the Tb-mode
to the global evolution of $f_3$ during the B-mode.
 At 112\,MHz we used a Discrete Fourier Transform to calculate $f_3$ for short sequences of pulses of the transitive modes.  In some cases we were forced to calculate $f_3$ on longer pulse sequences which included non-modulated Q-mode emission. For the LOFAR data we used mean values of the Tb-modes from Fig.~\ref{fig:F3}, with the errorbar corresponding to the standard deviation of $f_3$.

\begin{figure}
\centering
 \includegraphics[width=0.45\textwidth]{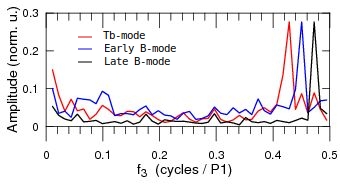}
 \caption{Examples of onpulse-integrated LRF spectra calculated for 91-pulse sequences containing the Tb-mode (150518, red line), early B-mode (260520, blue line), or late B-mode (221219, black line). Corresponding frequencies are 0.428, 0.449, and 0.472\,cycles/$\pone$.} 
 \label{fig:fig3}
\end{figure}

Examples of onpulse-integrated LRFS for three such longer 91-pulse sequences observed at 112\,MHz are shown in Fig.~\ref{fig:fig3}
for the Tb-mode (session 150518), early B-mode (session 260520), and late B-mode (221219). Each spectrum is  
normalized by the peak amplitude for clarity. The features corresponding to frequencies $f_3 = 0.428$, $0.449$, and $0.472$\,cycles/$\pone$, respectively, are narrow and well resolved\footnote{Note that for the Tb-mode some excess power is present at $f_3$ going up to 0.5\,cycles/$\pone$, corresponding to higher $f_3$ during the first half of the Tb-mode for session 150518 in Fig.~\ref{fig:F3}.}. It is obvious that $f_3$ gradually increases throughout the combined Tb- and B-modes.

\begin{figure}
\centering
 \includegraphics[width=0.45\textwidth]{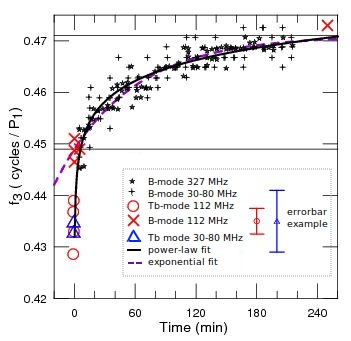}
 \caption{A compilation of observed $f_3$ values as a function of time since B-mode onset. Previously published $f_3$ measurements are shown by black stars (Arecibo, 327\,MHz) and black crosses (LOFAR, 30$-$80\,MHz). The 112-MHz measurements for the Tb- and B-modes from this work are shown with red circles and crosses. The $f_3$ value for 91 B-mode pulses right before a B-to-Q mode transition (session 221219) is plotted at an arbitrary time mark of 250\,min.  LOFAR Tb-mode $f_3$ measurements are shown with blue triangles.  The points around $t=180$\,min show typical $f_3$ errorbars for PRAO and LOFAR measurements from this work.
  Horizontal lines mark the range of validity of the exponential fit by \citet[][dashed magenta line]{Backus2011}. The black line shows power-law fit from \citet{Suleymanova2017}.
  The measurements for the Tb-mode are in a good accordance with a power law fit. } 
 \label{fig:fig4}
\end{figure}

A larger compilation of $f_3$ measurements is shown in Fig.~\ref{fig:fig4}. It comprises the $f_3$ values for the Tb- and B-modes from the current work, as well as LOFAR B-mode measurements from \citet{Bilous2018},  and Arecibo 327-MHz values from \citet{Rankin2006b}. As indicated in these studies and in \cite{Backus2011}, the frequency of the amplitude fluctuations varies with time according to the exponential law (dashed line), where $t$ is the time since mode onset (min):
\begin{equation}
f_3 = 0.471 - 0.022 \times e^{-t/73}.
\label{formula:n1}
\end{equation}
This equation limits $f_3$ to the range of $0.449-0.471$\,cycles/$\pone$. Numerous observations of \src\ at 112\,MHz showed 
that for roughly 2.5\% of B-mode pulse sequences $f_3<0.449$\,cycles/$\pone$. Thus,
\cite{Suleymanova2017} have proposed a better power-law parametrization that encompasses all B-mode $f_3$ measurements: 
\begin{equation}
f_3 = 0.439\times t^{0.0126}.
\label{formula:n2}
\end{equation}

In the Tb-modes, $f_3$ varies in the range of 0.428$-$0.439\,cycles/$\pone$ with a mean value of 0.434$\pm$0.004\,cycles/$\pone$ that is significantly lower than has ever been measured for the B-mode. Nevertheless, these values can be considered as compatible with the power law $f_3$-time dependence in the B-mode.  For both functions in Fig.~\ref{fig:fig4}, $t=0$ at B-mode onset. During the Tb-mode 
$t<0$, thus its $f_3$ values were not included in the power-law fitting procedure. The average value of $f_3$ at B-mode onset following the Tb-mode is 0.449$\pm$0.002\,cycles/$\pone$, as it was expected.

\subsection{Average pulse profile during transitive modes}
\begin{figure}
\centering
\includegraphics[width=0.25\textwidth]{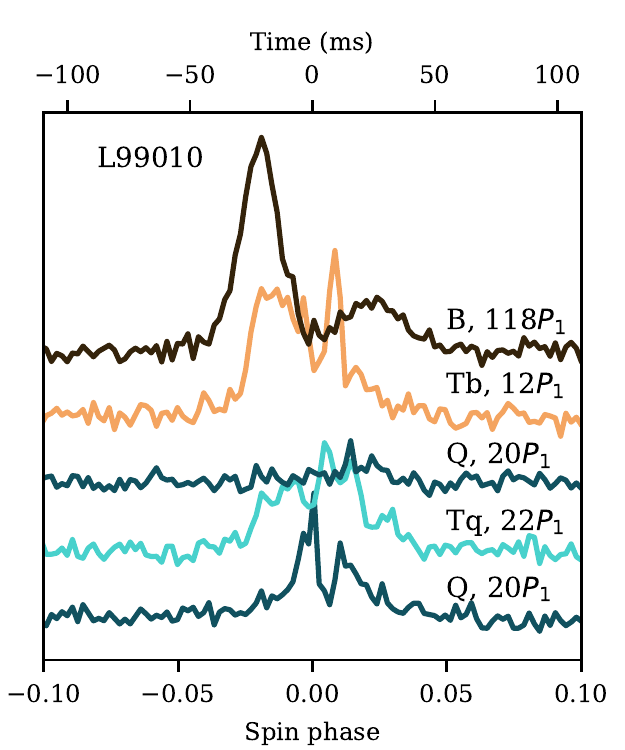}\includegraphics[width=0.25\textwidth]{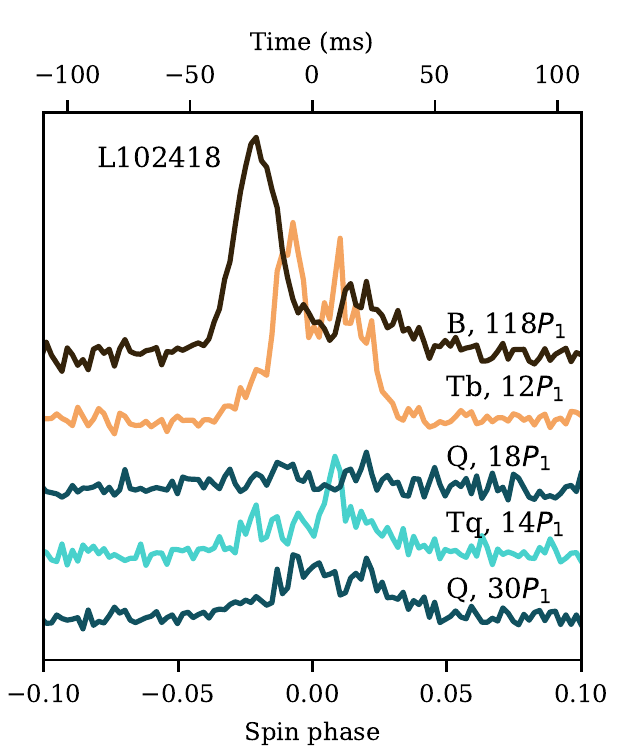}
  \caption{Average profiles integrated within the respective modes for two LOFAR sessions with Q-to-B transitions normalized by the standard
  deviation of signal outside the on-pulse window.}
\label{fig:prof_QB}
\end{figure}

Figure~\ref{fig:prof_QB} shows the average profiles for all four modes in two LOFAR sessions with Q-to-B transition. The shape 
of the average profile in transitive modes varies between the sessions and is dominated by the individual
strong pulses. Given the small number of modes and their relative shortness, it is hard to compare the average 
profiles in transitive modes with the regular B- and Q-modes, however it can be stated that neither the Tq- nor Tb-mode
exhibit two distinct separate components like the B-mode. 

\begin{figure}
\centering
 \includegraphics[width=0.45\textwidth]{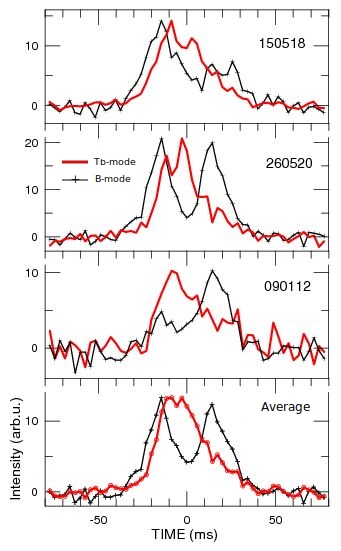}
  \caption{Average pulse shape of the Tb-mode (red line) and the B-mode (black line with crosses) for sessions (from top to bottom) 150518, 260520, and 090112. Profiles for different days were aligned by the leading peak of B-mode profile. The lowest panel shows profiles averaged over these three sessions.}
\label{fig:PRAO_BTb}
\end{figure}

PRAO observations offer larger mode samples and, also, longer instances of the Tb-mode. Figure~\ref{fig:PRAO_BTb} 
shows the average profiles integrated within B- and Tb-modes for the three PRAO sessions which exhibited 
prominent B-mode component peaks to facilitate alignment between sessions. 
The bottom panel presents integrated profiles for these three sessions, comprising 135 individual pulses for the Tb-mode and 390 pulses 
for the B-mode. It is evident that at 112\,MHz the Tb-mode has no separate components and the profile is asymmetric and skewed to earlier spin longitudes. 

\begin{figure}
\centering
 \includegraphics[width=0.42\textwidth]{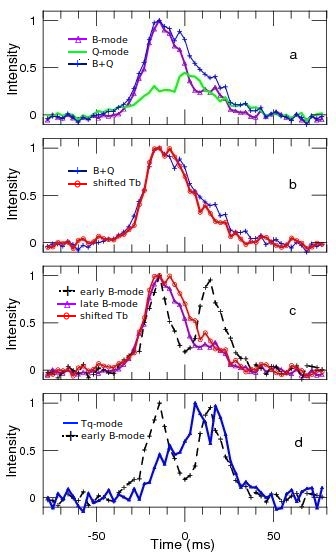}
  \caption{\textit{(a)}: mode-separated average profiles for session 221219, where the B-to-Q transition occurred in the middle of the observation. Violet and green lines correspond to the B- an Q-modes, respectively, and the thin blue line shows composite B+Q profile. \textit{(b)}: B+Q profile  (blue line) superposed on the average Tb-mode pulse profile (red line) shifted by 5.7\,ms towards earlier spin longitudes. \textit{(c)}: the B-mode pulse shape evolution starting from its onset  (session 260520, dotted line) up to cessation (session 221219, violet line), together with Tb-mode pulse (red line) shifted by 5.7\,ms towards earlier spin longitudes. \textit{(d)}: Tq-mode profile (blue line) accumulated from sessions 150518 and 051112. The B-mode profile on 260520 is given for comparison. Profiles for different days were aligned by the leading peak of the B-mode profile.
  }
\label{fig:PRAO_all4}
\end{figure}

Interestingly, the shape of the pulse profile in the averaged Tb-mode closely matches the composite Q+B profile recorded close to the B-to-Q 
transition (session 221219, Fig.~\ref{fig:PRAO_all4}a, b). In the session 221219, the B-to-Q switch happened in the middle of the observation.  For comparison, the average Tb-mode profile from Fig.~\ref{fig:PRAO_BTb} was shifted by 5.7\,ms toward earlier spin longitudes. This close resemblance of the pulse shapes 
could indicate that around the Q-to-B switch the regions responsible for the Q- and B-modes emit simultaneously and that their contribution is equal. This is possible only under the condition that the B- and Q-pulses are emitted from two independent regions in the pulsar magnetosphere. In the core/cone model \citep{Rankin1983}, the Q-  and B- modes in \src\ are associated with the core and conal  beams of emission which emit alternately. The discovery of  the transitive modes allow us to suggest that over some  short time these two independent regions may emit simultaneously.  

If the Tb-mode emission is indeed a superposition of the Q- and B-modes, then the 5.7\,ms delay is naturally explained by the evolution of the on-pulse window location throughout the B-mode. It is known that during the lifetime of each B-mode instance, the average pulse shifts towards later longitudes. The total shift varies from 4 to 6\,ms per B-mode duration \citep{Bilous2014, Suleymanova2014, Suleymanova2017}. Thus, the Tb-mode emission is a superposition of late B-mode and the Q-mode.
During the instantaneous Tb-to-B switch 
the emission window resets at earlier longitudes and the slow shift to the later longitudes continues during new B-mode instance.
Within this interpretation it is hard to explain though the B-Tb-B sequence of the 260520 session, where both B-mode sequences framing the Tb-mode exhibited the properties of the early B-mode. 

Observational evidence speaks against the Q-mode emission being produced by a core component in the 
core/cone model of  the pulsar radio beam. First of all, this evidence includes
the appearance of organized subpulse 
drift (Tq-mode) within the sequence of Q- mode pulses right before the Q-to-B switch and  the crude 
similarities between the shapes of the average profile in both modes. Also, at frequencies lower than 
100\,MHz Q-mode profiles exhibit conal components that separate progressively
\citep{Suleymanova1998,Bilous2014}.  It was shown that for \src\ the sightline makes only a grazing transverse of the polar cap \citep{Deshpande2000} and  the scenario in which the observer misses most of the core radiation because of this peripheral sightline was previously put forward to explain the relative faintness of \src’s radio emission during its bright X-ray mode \citep{Hermsen2013,Rankin2020}. 
Another scenario can be proposed in which core radio beam is missed completely and all modes observed in \src\ originate from conal beam shape variations. These variations are correlated with  \textbf{E}$\times$\textbf{B} drift behaviour of the localized spark  discharges.

Observations at 112\,MHz show that the width of the on-pulse window stays similar throughout all four modes (Fig.~\ref{fig:PRAO_all4}). It is considered to be determined by the size of the polar cap with the radius $R_\mathrm{out}$. Within the framework of the traditional core/cone model of pulsar radio emission \citep[e.g.][and other papers in this series]{Rankin1983}, the width of the conal components in the average pulse profile is set by the thickness of the pair-producing ring in the polar gap: $W_\mathrm{ring}=R_\mathrm{out}-R_\mathrm{inn}$, where $R_\mathrm{inn}$ is the inner radius of the ring. We suggest that the transitions between modes are accompanied by variations in the inner radius. This can explain the gradual increasing of the component’s width over the B-mode lifetime and the appearance of drifting subpulses at the central longitudes in the Tb and Tq-modes. At B-mode onset the two components of the average pulse profile are well separated with intensity in the saddle region close to zero below 100 MHz.  The width of the components is at its minimum and grows with B-mode age  \citep{Bilous2014}. For the leading component 112\,MHz, the width can increase by as much as 40\% \citep{Suleymanova2014}. This effect is well demonstrated in Fig.~\ref{fig:PRAO_all4}c, where the profiles of the average pulse for the onset and the end of the B-mode are shown. The width of the first component at the level of half maximum increases from 15 to 24\,ms, i.e. by 60\%. The same panel shows the  profile of the average Tb-mode with the width of 30\,ms. The total change in profile width between the Tb- and early B-mode is 100\%. With such a significant increase in the width of $W_\mathrm{ring}$, central longitudes become  illuminated for our line of sight and we observe drifting subpulses close to fiducial longitude in transitive modes.

Figure~\ref{fig:PRAO_all4}d shows the Tq-mode profile obtained at 112\,MHz after averaging 20 pulses using sessions 150518 and 051112. It is worth noting that because of the short duration of the mode its average profile is distorted by bright pulses. However, comparing the Tq-mode to the Q-mode profile (Fig.~\ref{fig:PRAO_all4}a) reveals similarity in profile shape as far as they are both skewed to the trailing side (unlike B-mode). Similar behaviour is observed in LOFAR data (Fig.~\ref{fig:prof_QB}).

\section{Summary}

In this work we report on the discovery of a transitive process which takes place for about a minute around a Q-to-B-mode switch. This process consists of two stages which we labelled Tb- and Tq- modes.
  
\subsection{Tb-mode and its comparison to the B-mode.}

\begin{enumerate}
 \item This mode  precedes the B-mode onset by 40$\pm$25\,s. The duration of the Tb-modes exhibit larger variations from one instance to another, ranging between $12\pone$\ and $66\pone$. 
 
 \item Similarly to B-mode, the individual subpulses drift from the leading to trailing edge of the onpulse window, however the drift extends now through the whole onpulse window, including central spin longitudes.
 
 \item The amplitude modulation frequency in the Tb-mode varies in the range of 0.428$-$0.439\,cycles/$\pone$ with the mean value of 0.434$\pm$0.005\,cycles/$\pone$ that is significantly lower than $f_3$ values measured for the B-mode. Nevertheless, these values are in accordance with the power-law $f_3$ evolution throughout B-mode.  
 
 \item The average pulse profile in the Tb-mode is skewed to the leading edge, similar to the B-mode, however it does not have resolved conal components.
 
 \item Sometimes the Tb-mode happens within an early B-mode instance.
\end{enumerate}

\subsection{Tq-mode and its comparison to the Q-mode.}

\begin{enumerate}
 \item The Tq-mode occurs within the late Q-mode and, unlike the latter, has a specific subpulse drift pattern. 
 This mode  precedes the Tb-mode onset by 34$\pm$7\,s and lasts for about $10-22\pone$. In one session no Tq-mode was recorded, however the S/N of the recording was quite low.
 
 \item As in the Tb- and B-modes, subpulses in the Tq-mode drift towards the trailing edge of the onpulse window. The drift frequency $f_3$ varies from 0.149 to 0.312\,cycles/$\pone$ from one observation to another, with sometimes large variation of $f_3$ within the mode. 

 \item The average pulse profile in the Tq-mode has a tendency to be skewed to the trailing edge which is characteristic of the Q-mode.
 
\end{enumerate}

\section{Discussion}

Short-lived mode sequences have been previously observed from several pulsars, for example  PSR B1859+17 \citep{Rajwade2021}, 
PSR~J1326$-$6700 \citep{Wen2020}, PSR J1909$-$3744 \citep{Miles2021}, and others. The peculiarity of \src's transitive modes lies in the 
fact that they usually appear in a 1-minute interval around the switch between the two main hours-long modes, the Q- and B-mode. Both 
modes show similarities to the main modes and are considered as submodes, which is reflected in their designation.

\cite{Gil2000} argued that subpulse drift is observed when a quasi-central spark is formed at the local pole of 
a sunspot-like surface magnetic field. This fixed spark prevents other sparks from moving towards the pole, restricting their motion
to slow drift around the pole. Although little is known about \src's small-scale magnetic field configuration, we can qualitatively 
apply this model to \src's mode sequences. During the Q-mode no central spark is formed, sparks are quickly moving towards magnetic 
pole and subpulses are observed at random onpulse longitudes. At the end of the Q-mode the anchor spark appears, causing 
Tq-mode sequences. The carousel rotates relatively slowly and exhibits a lot of variation from one mode instance to another, as well as within the Tq-mode, possibly reflecting fast-paced changes in the spatial or temporal variations of accelerating potential in the polar gap.
Subpulses are seen in the central region of the on-pulse window, indicating a larger width of the carousel. Subpulses briefly disappear (except for may be a few pulses) during the null-like period between the Tq- and Tb-modes. During the Tb-mode the carousel width remains large, but then gradually decreases throughout the B-mode. The carousel rotation rate evolves in a similar manner, more rapidly during the Tb- and early B-mode 
and more slowly at the end. During the reverse B-to-Q transition the anchor spark disappears and the drift ceases. 

Drifting subpulses are traditionally used as a voltmeter to measure the gradient of accelerating potential across the polar cap \citep{vanLeeuwen2012}. Assuming a small degree of aliasing in the B-, Tb- and Tq-modes, the large and fast fractional change of $\pthr$ in the Tb- and Tq-modes (comparing to the more gradual smaller change over the course of B-mode) means some rapid and powerful magnetospheric current rearrangements. 

The timescales of mode switching for \src\ and other mode-switching pulsars are much larger than characteristic time scales of force-free magnetospheres ($\sim\pone$). It has been  conjectured that mode switching are a manifestation of meta-stable magnetospheric states, produced by non-linear interaction between the neutron star magnetic fields, polar cap cascades and current sheets \citep{Timokhin2010}. In particular, these quasi-stable states can have different open fieldline zone sizes and/or different current density distributions within the open field line zone. If switches between the B- and Q-mode are accompanied by a shrinking or expanding of the open field line zone (``polar cap'', for dipole external magnetic field), then, assuming that the radio emission comes from the last open field lines one may infer the polar cap radius from the width of profiles in the B- and Q-modes. Taking 0.06 of spin phase for the former and 0.05 for the latter, for the plausible ranges of the inclination angle and LOS impact angle \citep{Bilous2018}, one may infer a 5\% shrinking of the polar cap radius in the Q-mode. According to \citet{Timokhin2010} such change would correspond to a 20\% change in spin-down rate between the B- and Q-modes. This change is not possible to detect directly for the $3\times10^{-15}$\,s/s spin down rate of \src\ considering that mode switching happens on a timescale of hours. However, if the relative fraction of Q- and B-modes change slowly over time, this would have influence on the overall spin-down rate. A similar effect was detected for PSR B1828$-$11, where a variable rate of mode transitions were directly related to the spin-down changes \citep{Stairs2019}. 

Our crude estimate above implies that the size of polar cap (PC) is larger in the B-mode. \citet{Rigoselli2018} performed direct fitting of the mode-separated thermal X-ray spectra and lightcurves. Generally, the Q-mode has larger polar cap size and larger temperature than B-mode,  however the error contours do permit 5\% larger PCs radius in B-mode. For this, Q-mode PCs should have about a 20\% higher temperature. 

On the other hand, \citet{Szary2015} proposed that mode switching reflects switching between two kinds of partially screened gap, with different pair production and a gap screening mechanism. 
For both kinds the PC temperature should be roughly the same in order to maintain the thermostat mechanism of the partially screened gap. This also does not contradict the results of \citet{Rigoselli2018}, although in this case the size of PC in the Q-mode should be larger. We must note that the heating patterns are generally poorly known and the PC is almost certainly not illuminated uniformly. Radio emission may also be a poor indicator of the PC size, coming from the same field lines in both modes but shifting in height. 

So far both transitive modes exhibited quite large variations of properties from one session to another. In order to deepen our understanding of the carousel formation, more mode instances should be explored, preferably in a wide frequency range.

\begin{acknowledgements}
AB acknowledges the support from
the European Research Council under the European Union's Seventh Framework Programme
(FP/2007-2013)/ERC Grant Agreement No. 617199 (`ALERT') and
Vici research programme `ARGO' with project number
639.043.815, financed by the Dutch Research Council (NWO).
\end{acknowledgements}

\bibliographystyle{aa} 
\bibliography{0943_bibliography}

\end{document}